\documentclass[aps,showpacs,superscriptaddress,preprint]{revtex4-2}
\usepackage{latexsym,graphicx,epsfig,psfrag}

\usepackage{epstopdf}
\usepackage{amssymb,amsmath,amsxtra,amsfonts}
\usepackage{bm}

\usepackage{longtable}
\usepackage{multirow}
\usepackage{booktabs}
\usepackage{array}
\usepackage{wrapfig}
\usepackage{color}
\usepackage{titlesec}

\titleformat{\section}{\large\bfseries}{\thesection}{1em}{}

\newcommand{\bea}{\begin{eqnarray}}
\newcommand{\ena}{\end{eqnarray}}
\newcommand{\be}{\begin{equation}}
\newcommand{\en}{\end{equation}}
\newcommand{\nn}{\nonumber\\}
\newcommand{\ed}{\end{document}} 
\newcommand{\Tr}{\mbox{\rm{tr}}}

\newcommand{\slp}{p\kern-5pt/}

\begin{document}

\title{Study of the semileptonic decays \boldmath{$\Upsilon(1S)\to B_{(c)}\ell\bar{\nu}_\ell$}} 

\author{C.~T.~Tran}
\email{thangtc@hcmute.edu.vn}
\affiliation{
	\hbox{Department of Physics, HCMC University of Technology and Education, }\\
	Vo Van Ngan 1, 700000 Ho Chi Minh City, Vietnam}

\author{M. A. Ivanov}
\email{ivanovm@theor.jinr.ru}
\affiliation{Bogoliubov Laboratory of Theoretical Physics, 
Joint Institute for Nuclear Research, 141980 Dubna, Russia}

\author{P.~Santorelli}
\email{Pietro.Santorelli@na.infn.it}
\affiliation{
	Dipartimento di Fisica ``E.~Pancini'', Universit\`a di Napoli Federico II,
	\hbox{ Complesso Universitario di Monte S.~Angelo,
		Via Cintia, Edificio 6, 80126 Napoli, Italy}}
\affiliation{
	\hbox{Istituto Nazionale di Fisica Nucleare, 
		Sezione di Napoli, 80126 Napoli, Italy}}

\author{H.~C.~Tran}
\email{catth@hcmute.edu.vn}
\affiliation{
	\hbox{Department of Physics, HCMC University of Technology and Education, }\\
	Vo Van Ngan 1, 700000 Ho Chi Minh City, Vietnam}

\begin{abstract}
We study the exclusive semileptonic decays $\Upsilon(1S)\to B_{(c)}\ell\bar{\nu}_\ell$, where $\ell = e,\mu,\tau$. The relevant hadronic form factors are calculated using the Covariant Confined Quark Model developed previously by our group. We predict the branching fractions $\mathcal{B}(\Upsilon(1S)\to B_{(c)}\ell\bar{\nu}_\ell)$ to be of the order of $10^{-13}$ and $10^{-10}$ for the case of $B$ and $B_c$, respectively. Our predictions agree well with other theoretical calculations. We also consider the effects of possible New Physics in the case of $\Upsilon(1S)\to B_c\tau\bar{\nu}_\tau$. We show that the branching fraction of this decay can be enhanced by an order of magnitude using constraints from the $B\to D^{(*)}\ell\bar{\nu}_\ell$ and $B_c\to J/\psi\ell \bar{\nu}_\ell$ experimental data.    
 
\end{abstract}


\maketitle
\newpage

\section{Introduction}
\label{sec:intro}
Low-lying quarkonia systems such as $\Upsilon(1S)$ mostly decay through intermediate gluons or photons produced by the parent $\bar{q}q$ pair annihilation. As a result, strong and radiative decays of $\Upsilon(1S)$ have been widely studied, both theoretically and experimentally. Meanwhile, weak decays of $\Upsilon(1S)$ have attracted less attention. Thanks to the significant progress achieved in recent years in the improvement of luminosity of colliders, a large amount of rare weak decays have been observed. In particular, the rare semileptonic decay of the charmonium $J/\psi\to D\ell^+\nu_{\ell}$, ($\ell = e,\mu$), was considered one of the main research topics at BESIII experiment~\cite{BESIII:2020nme}. In 2021, BESIII reported a search for the decay $J/\psi\to D e^+\nu_e$ based on a sample of $10.1\times 10^9$ $J/\psi$ events~\cite{BESIII:2021mnd}. The result placed an upper limit of the branching fraction to be $\mathcal{B}(J/\psi\to D e^+\nu_e+c.c.) < 7.1 \times 10^{-8}$ at 90~\% confidence level (CL). It is worth mentioning that this upper limit was improved by a factor of 170 compared to the previous one~\cite{BES:2006mls}. In 2023, using the same $J/\psi$ event sample, BESIII searched for the semimuonic channel for the first time and found an upper limit of $\mathcal{B}(J/\psi\to D \mu^+\nu_\mu+c.c.) < 5.6 \times 10^{-7}$ at 90~\% CL~\cite{BESIII:2023fqz}. These upper limits are still much larger than the Standard Model (SM) predictions, which are of order of $10^{-11}$~\cite{Wang:2007ys,Shen:2008zzb,Dhir:2009rb,Ivanov:2015woa,Wang:2016dkd}. Nevertheless, the experimental data implied constraints on several New Physics models which can enhance the branching fractions to the order of $10^{-5}$~\cite{Datta:1998yq}. In the light of the extensive search for rare charmonium decays, it is reasonable to explore similar decays of the bottomonium $\Upsilon(1S)$.    

The semileptonic decays $\Upsilon(1S)\to B_{(c)}\ell\bar{\nu}_\ell$, where $\ell = e,\mu,\tau,$ have been investigated in several theoretical studies. However, there are very few of them. Besides, the existing predictions still differ. First calculation of the decays $\Upsilon(1S)\to B_c\ell\bar{\nu}_\ell$ was carried out by Dhir, Verma, and Sharma~\cite{Dhir:2009rb} in the framework of the Bauer-Stech-Wirbel model. They obtained $\mathcal{B}(\Upsilon(1S)\to B_c e \bar{\nu}_e)=\big(1.70^{+0.03}_{-0.02}\big)\times 10^{-10}$ and $\mathcal{B}(\Upsilon(1S)\to B_c \tau \bar{\nu}_\tau)=\big(2.9^{+0.05}_{-0.02}\big)\times 10^{-11}$. In this paper, the authors only considered the $\Upsilon(1S)\to B_c$ transition. In 2017, Wang \emph{et~al.} calculated the decays $\Upsilon(1S)\to B^{(*)}_{(c)}\ell\bar{\nu}_\ell$ using the Bethe-Salpeter method~\cite{Wang:2016dkd}. The results for the $\Upsilon(1S)\to B_c$ case read $\mathcal{B}(\Upsilon(1S)\to B_c e \bar{\nu}_e)=\big(1.37^{+0.22}_{-0.19}\big)\times 10^{-10}$ and $\mathcal{B}(\Upsilon(1S)\to B_c \tau \bar{\nu}_\tau)=\big(4.17^{+0.58}_{-0.52}\big)\times 10^{-11}$. The results of the two studies above only marginally agree with each other. Let us now consider the ratio of branching fractions, namely, $R(\Upsilon(1S)\to B_c)=\mathcal{B}(\Upsilon(1S)\to B_c \tau \bar{\nu}_\tau)/\mathcal{B}(\Upsilon(1S)\to B_c e \bar{\nu}_e)$. Based on the branching fractions given above, we estimate $R(\Upsilon(1S)\to B_c)$ to be about $0.17\pm 0.01$ (Dhir \emph{et~al.}) and $0.30\pm 0.09$ (Wang \emph{et~al.}). The results imply a tension at $1.5~\sigma$ between the two studies. 
%%%%% begin adding
Moreover, in 2016, Chang \emph{et~al.} considered the decays $\Upsilon(nS)\to B_c\ell\bar{\nu}_\ell$ ($n=1,2,3$) based on nonrelativistic QCD (NRQCD)~\cite{Chang:2016gyw}. They found $R(\Upsilon(1S)\to B_c)= 0.24^{+0.02}_{-0.01}$, which agrees with Wang \emph{et~al.} but disagrees with Dhir \emph{et~al.}. 
%%%% end adding
It is, therefore, necessary to provide more  theoretical predictions for the decays.  

There is another interesting aspect of the decay $\Upsilon(1S)\to B_c\ell\bar{\nu}_\ell$. At quark level, it is induced by the transition $b\to c\ell\bar{\nu}_\ell$. For more than a decade, tensions between experimental data and the Standard Model predictions for the ratios of branching fractions $R_D=\mathcal{B}(B^0\to D \tau \bar{\nu}_\tau)/\mathcal{B}(B^0\to D e \bar{\nu}_e)$ and $R_{D^*}=\mathcal{B}(B^0\to D^* \tau \bar{\nu}_\tau)/\mathcal{B}(B^0\to D^* e \bar{\nu}_e)$ have never disappeared. It is well known in the community as ``the $R_{D^{(*)}}$ puzzle'', which hints possible violation of lepton flavor universality (LFU) and motivates a huge search for New Physics in the semileptonic decays $B^0\to D^{(*)} \ell \bar{\nu}_\ell$ (see, e.g., \cite{Groote:2021ayy,Ivanov:2020iad} and references therein). The decay $\Upsilon(1S)\to B_c\ell\bar{\nu}_\ell$ is therefore a reasonable candidate to probe possible New Physics beyond the SM.

Weak decays of hadrons such as $\Upsilon(1S)\to B_{(c)}\ell\bar{\nu}_\ell$ are characterized by the interplay of strong and weak interactions. While the structure of the weak interaction in semileptonic decays is well established, the strong interaction in the hadronic transitions $\Upsilon(1S)\to B_{(c)}$ can only be calculated using nonperturbative methods. Hadronic transitions are often parametrized by invariant form factors. In this paper, hadronic form factors of the semileptonic decays of $\Upsilon(1S)$ are calculated in the framework of the covariant confined quark model (CCQM) developed previously by our group. One of the advantages of our model is the ability to calculate the form factors in the whole physical range of momentum transfer without any extrapolation.

The rest of the paper is organized as follows. In Sec.~\ref{sec:formalism} we present the relevant theoretical formalism for the calculation of the semileptonic decays $\Upsilon(1S)\to B_{(c)}\ell\bar{\nu}_\ell$. In Sec.~\ref{sec:model} we briefly introduce the CCQM and demonstrate the calculation of the hadronic form factors in our model. We then present our numerical results in Sec.~\ref{sec:result} and conclude in Sec.~\ref{sec:sum}.

%-----------------------------------------------------------------
\section{Formalism}
\label{sec:formalism}
\begin{figure}[t]
	\begin{center}
		%\hspace*{-0.5cm}
		\begin{tabular}{c} 
			\includegraphics[width=0.5\textwidth]{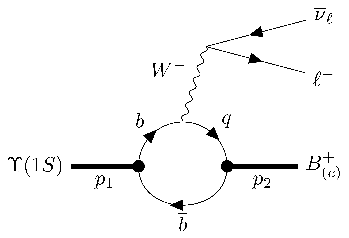}
		\end{tabular}
	\end{center}
	\caption{\label{fig:semilept}
		Feynman diagram for semileptonic decays $\Upsilon(1S)\to B_{(c)}\ell\bar{\nu}_\ell$.}
\end{figure}

In the CCQM the semileptonic decays $\Upsilon(1S)\to B_{(c)}\ell\bar{\nu}_\ell$ are described by the Feynman diagram in Fig.~\ref{fig:semilept}. The effective Hamiltonian for the semileptonic decays $\Upsilon(1S)\to B_{(c)}\ell\bar{\nu}_\ell$ is given by
\begin{equation}
\mathcal{H}_{\textrm{eff}} (b \to q \ell \bar{\nu}_{\ell}) = 
\frac{G_F}{\sqrt{2}} V_{qb} 
\left[ \bar{q} O^\mu  b \right] 
\left[ \bar{\ell} O_\mu  \nu_{\ell} \right],
\label{eq:Hamiltonian}
\end{equation}
where $q = (u, c)$, $G_F$ is the Fermi constant, $V_{qb}$ is the Cabibbo-Kobayashi-Maskawa matrix element, and $O^\mu=\gamma^\mu(1-\gamma_5)$ is 
the weak Dirac matrix with left chirality. The invariant matrix element of the decays is written as
\begin{equation}
\mathcal{M}(\Upsilon(1S)\to B_{(c)} \ell \bar\nu_\ell )  =  
\frac{G_F}{\sqrt{2}} V_{qb}
\left\langle B_{(c)}|\bar{q}O^\mu b|\Upsilon(1S)\right\rangle
\bar\ell O_\mu \nu_\ell.
\label{eq:M}
\end{equation}
%%%
%\vspace*{-0.5cm}

The squared matrix element can be written as a product of the hadronic tensor $H_{\mu\nu}$ and leptonic tensor $L^{\mu\nu}$:
\begin{equation}
	|\mathcal{M}|^2 = \frac{G_F^2}{2} H_{\mu\nu} L^{\mu\nu}.
\end{equation}
The leptonic tensor for the process
$W^-_{\rm off-shell}\to \ell^-\bar \nu_\ell$ 
$\left( W^+_{\rm off-shell}\to \ell^+ \nu_\ell \right)$  
is given  by \cite{Gutsche:2015mxa}
\begin{eqnarray}
L^{\mu\nu} &=& \left\{\begin{array}{lr}
	\Tr\Big[ (\slp_\ell + m_\ell) O^\mu \slp_{\nu_\ell} O^\nu\Big]
	& \qquad\text{for}\qquad  W^-_{\rm off-shell}\to \ell^-\bar\nu_\ell 
	\\[1.2ex]
	\Tr\Big[ (\slp_\ell - m_\ell) O^\nu \slp_{\nu_\ell} O^\mu\Big]
	& \qquad\text{for}\qquad  W^+_{\rm off-shell}\to \ell^+ \nu_\ell 
\end{array}\right.
\nn[1.2ex]
&=&
8 \left( 
p_\ell^\mu p_{\nu_\ell}^\nu  + p_\ell^\nu p_{\nu_\ell}^\mu 
- p_{\ell}\cdot p_{\nu_\ell}g^{\mu\nu}
\pm  i \varepsilon^{\mu \nu \alpha \beta} p_{\ell\alpha} p_{\nu_\ell\beta}
\right),
\label{eq:lept_tensor}
\end{eqnarray}
where the upper/lower sign 
refers to the two $(\ell^-\bar\nu_\ell)/(\ell^+\nu_\ell)$ configurations.
The sign change is due to the parity violating part of 
the lepton tensors. In our case we have to use the upper sign
in Eq.~(\ref{eq:lept_tensor}).

The hadronic matrix element in Eq.~(\ref{eq:M}) is often parametrized as a linear combination of Lorentz structures multiplied by scalar functions, namely, invariant form factors which depend on the momentum transfer squared. For the $V \to P$  transition one has
\begin{align}
	&\langle B_{(c)} (p_2) \left|\bar{q} O_{\mu} b \right| 
	\Upsilon(1S) (\epsilon_1 , p_1)\rangle \equiv \epsilon^\alpha_1 T_{\mu\alpha}
	\nn
	&=\frac{\epsilon_1^{\nu}}{m_1 + m_2}
	[ - g_{\mu\nu}pqA_0(q^2) + p_{\mu}p_{\nu}A_+(q^2)+q_{\mu}p_{\nu}
	A_-(q^2) + i \varepsilon_{\mu\nu\alpha\beta}p^{\alpha}q^{\beta}V(q^2)],
	\label{eq:FFpseudoscalar}
\end{align}
where $q=p_1-p_2$, $p=p_1+p_2$, $m_1 = m_{\Upsilon(1S)}$, $m_2 = m_{B_{(c)}}$, and $\epsilon_1$ is the polarization vector of $\Upsilon(1S)$, so that $\epsilon_1^\dagger\cdot p_1 = 0$. The particles are on-shell, i.e., $p_1^2=m_1^2=m^2_{\Upsilon(1S)}$ and $p_2^2=m_2^2=m^2_{B_{(c)}}$. The form factors $A_0(q^2)$, $A_{\pm}(q^2)$, and $V(q^2)$ will be calculated later in our model. In terms of the invariant form factors, the hadronic tensor reads
\begin{equation}
	H_{\mu\nu} = 
	T^{VP}_{\mu\alpha}\left(-g^{\alpha\alpha'}+\frac{p_1^\alpha p_1^{\alpha'}}{m_1^2}\right)
	T^{VP\dagger}_{\nu\alpha'}, 
\end{equation}
where
\begin{equation}
	T^{VP}_{\mu\alpha}=\frac{1}{m_1 + m_2}
	\left[- g_{\mu\alpha}pqA_0(q^2) + p_{\mu}p_{\alpha}A_+(q^2)+q_{\mu}p_{\alpha}
	A_-(q^2) + i \varepsilon_{\mu\alpha\gamma\delta}p^{\gamma}q^{\delta}V(q^2)\right] .
\end{equation}

Finally, by summing up the vector polarizations, one obtains the decay width
\begin{equation}
\Gamma\left (\Upsilon(1S) \to B_{(c)}\ell \bar{\nu}_{\ell} \right)
= \frac{G_F^2}{(2\pi)^3}\frac{|V_{qb}|^2}{64m_1^3}
\int\limits_{m^2_{\ell}}^{(m_1-m_2)^2}\!\!\!\! dq^2
\int\limits_{s_1^-}^{s_1^+}\!\! ds_1
\frac13 H_{\mu\nu} L^{\mu\nu},
\label{eq:rate}
\end{equation}
where $m_1=m_{\Upsilon(1S)}$, $m_2=m_{B_{(c)}}$, and $s_1 =(p_{B_{(c)}}+p_{\ell})^2$. 
The upper and lower bounds of $s_1$ are given by
\begin{equation}
s_1^{\pm}=
m_2^2+m_{\ell}^2-\frac{1}{2q^2}
\left[(q^2-m_1^2+m_2^2)(q^2+m_{\ell}^2)
\mp\lambda^{1/2}(q^2,m_1^2,m_2^2)\lambda^{1/2}(q^2,m_{\ell}^2,0)\right],
\end{equation}
where $\lambda(x,y,z) \equiv x^2+y^2+z^2-2(xy+yz+zx)$ is 
the K{\"a}ll{\'e}n function. 
%-----------------------------------------------------------------

\section{Form factors in the Covariant Confined Quark Model}
\label{sec:model}
\subsection{CCQM in a nutshell}
The CCQM has been developed for about three decades as a tool for hadronic calculation. It has been successfully employed to explore various decays of not only mesons and baryons, but also tetraquarks, pentaquarks, and other multiquark states~\cite{Dubnicka:2010kz, Dubnicka:2011mm, Gutsche:2014zna, Goerke:2016hxf, Ivanov:2023wir, Groote:2021ayy, Dubnicka:2020yxy, Dubnicka:2020xoh, Gutsche:2017twh, Gutsche:2017hux, Goerke:2017svb}. The model has been introduced in great details along the way in many studies by our group, for instance, Refs.~\cite{Branz:2009cd, Ivanov:2006ni,Tran:2023hrn}. We only list here the main features of the model for completeness, and also, to keep the text short and focus more on the new results. 

%%%% begin adding
The  CCQM is based on an effective interaction Lagrangian describing the
coupling of hadrons to their constituent quarks.
The coupling of a meson $M(q_1 \bar q_2)$ to its constituent 
quarks $q_1$ and $\bar q_2$ is described by the nonlocal Lagrangian (see, e.g., Ref.~\cite{Branz:2009cd}) 
\be
\label{eq:lag}
{\cal L}_{\rm int}(x) = g_M M (x) \, 
\int\!\! dx_1 \!\! \int\!\! dx_2 
F_M(x,x_1,x_2) \,
\bar q_1(x_1) \, \Gamma_M \, q_2(x_2)  \, + \, {\rm H.c.}  
\en 
Here, $g_M$ is the meson-quark coupling constant, $\Gamma_M$  the relevant Dirac matrix (or a string of Dirac matrices)
chosen appropriately to describe the spin quantum numbers of the 
meson field $M(x)$.
The vertex function $F_M(x,x_1,x_2)$ characterizes 
the finite size of the meson.  To satisfy translational invariance the 
vertex function has to obey the identity 
$F_M(x+a,x_1+a,x_2+a) \, = \, F_M(x,x_1,x_2) $
for any given four-vector $a$. 
In what follows we adopt a specific form for the vertex function which 
satisfies the above translation invariance relation. One has 
\begin{equation}
	F_M(x;x_1,x_2) = \delta^{(4)}(x-\omega_1 x_1-\omega_2 x_2)\Phi_M[(x_1-x_2)^2],
\end{equation} 
where $\Phi_M$ is a correlation function of the two constituent quarks 
with masses $m_1$ and $m_2$. 
The variable $w_i$ is defined by $w_i=m_i/(m_1+m_2)$ so that 
$w_1+w_2=1$. 
In principle, the Fourier transform of the correlation function, which we
denote by $\Phi_M(-p^2)$, can be calculated from the solutions
of the Bethe-Salpeter equation for the meson bound  
states~\cite{Roberts:2000aa}. 

In Ref.~\cite{Anikin:1995cf} it was found that, using various forms for the vertex function, the basic hadron observables are insensitive to the details of
the functional form of the hadron-quark vertex function. Therefore, we use this observation as a guiding principle and choose a simple Gaussian form for the Fourier transform of the correlation function $\Phi_{M}(-p^2)$. The minus sign in the argument is chosen to emphasize that we are working in Minkowski space. One has 
\begin{equation}
\label{eq:vertexf}
\Phi_{M}(-p^2) = \exp( p^2/\Lambda_M^2), 
\end{equation}
where the parameter $\Lambda_M$ characterizes the size of the meson. Since $p^2$ turns into $-p_E^2$ in  Euclidean space, 
the form~(\ref{eq:vertexf}) has the appropriate falloff
behavior in the Euclidean region. We stress again that any choice for 
$\Phi_M$ is appropriate as long as it falls off sufficiently fast
in the ultraviolet region of Euclidean space to render the Feynman-diagram
ultraviolet finite. 

Recently, we have also considered a more complicated structure of the vertex functions. In Ref.~\cite{Dubnicka:2024geu}, we introduced the vertex functions as Gaussian multiplying some polynomial to describe the radial excitations of the charmonium and bottomonium as follows:
\[
\Phi_{n}(-k^{2})=\Big(1+\sum\limits _{m=1}^{n}c_{m+n-1}s_{n}^{m}k^{2m}\Big)\Phi_{V}(-k^{2}),
\]
where $\Phi_{V}(-k^{2})=\exp(k^2/\Lambda^2_V)$.
Here we have assumed that different radial excitations cannot pass into each
other through a quark loop. We called this assumption the orthogonality
condition. This condition allowed us to determine numerical values of
the coefficients $c_{m+n-1}$ in the allowed region of $p^{2}$ for a
certain quarkonium spectrum.

%%%% end adding

The quark-meson coupling is obtained using the compositeness 
condition~\cite{Salam:1962ap,Weinberg:1962hj}
\begin{equation}
Z_M = 1 - \Pi^\prime_M(m^2_M) = 0,
\label{eq:compositeness}
\end{equation}
where $Z_M$ is the wave function renormalization constant of the meson $M$ 
and $\Pi'_M$ is the derivative of the meson mass function.
%%%%%%
\begin{figure}[ht]
	\includegraphics[width=0.40\textwidth]{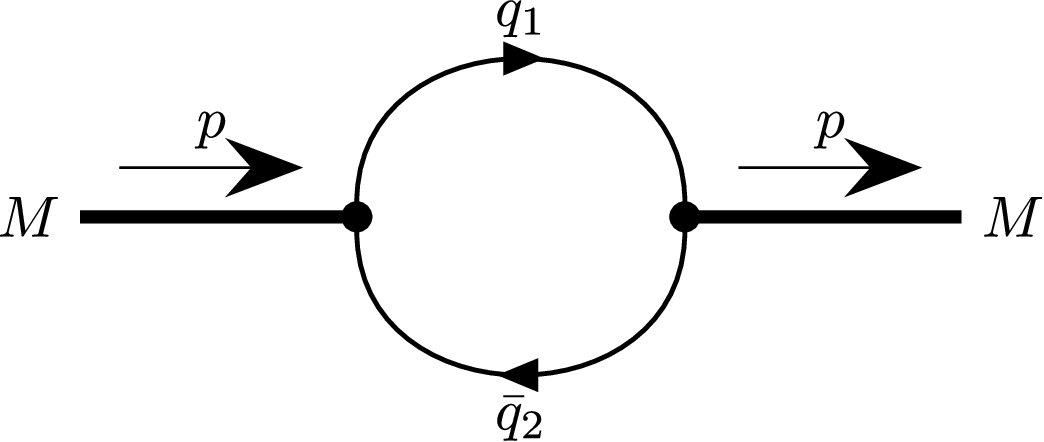}
	\caption{One-loop self-energy diagram for a meson.}
	\label{fig:mass}
\end{figure}
%%%%%%%%%%

The meson mass function 
in Eq.~(\ref{eq:compositeness}) is defined by the Feynman diagram shown
in Fig.~\ref{fig:mass} and has the following form: 
\begin{eqnarray}
\Pi_P(p) &=& 3g_P^2 \int\!\! \frac{dk}{(2\pi)^4i}\,\widetilde\Phi^2_P \left(-k^2\right)
\Tr\left[ S_1(k+w_1p)\gamma^5 S_2(k-w_2p)\gamma^5 \right],\\
\label{eq:Pmass-Pseudoscalar}
\Pi_V(p) &=& g_V^2 \left[g^{\mu\nu} - \frac{p^{\mu}p^{\nu}}{p^2}\right] 
\int\!\! \frac{dk}{(2\pi)^4i}\,\widetilde\Phi^2_V \left(-k^2\right)
\Tr\left[ S_1(k+w_1p)\gamma_{\mu} S_2(k-w_2p)\gamma_{\nu} \right],
\end{eqnarray}
where 
\begin{equation}
S_i(k) = \frac{1}{m_{q_i} - \not\! k - i\epsilon}= \frac{m_{q_i}+\not\! k}{m^2_{q_i} - k^2 - i\epsilon}
\label{eq:prop}
\end{equation}
is the quark propagator.

The CCQM has several free parameters including the constituent quark masses $m_q$, the hadron size parameters $\Lambda_H$, and a universal cutoff parameter $\lambda$ which guarantees the confinement of constituent quarks inside hadrons. These parameters are obtained by fitting to available experimental data and/or Lattice QCD. Once they are fixed, the CCQM can be used to calculate hadronic quantities in a straight-forward manner. The parameters relevant to this study are collected in Table~\ref{tab:para}.
%%%%%%%%%%%%
\begin{table}[ht]
	\caption{Quark masses, meson size parameters, and infrared cutoff parameter (all in GeV).}\label{tab:para}
	\renewcommand{\arraystretch}{0.7}
	\begin{ruledtabular}
		\begin{tabular}{cccccccc}
			$m_{u/d}$ & $m_s$ &  $m_c$ &  $m_b$ & $\Lambda_B$ &	$\Lambda_{B_c}$ & $\Lambda_{\Upsilon(1S)}$ & $\lambda$ \\
			\hline
			0.241 & 0.428 & 1.67 & 5.04 & 1.96 &	2.73 & 4.03 & 0.181
		\end{tabular}
	\end{ruledtabular}
\end{table}
%%%%%%%%%

\subsection{Hadronic matrix element and form factors}
In the CCQM the hadronic matrix element of the semileptonic decays $\Upsilon(1S)\to B_{(c)}\ell\bar{\nu}_\ell$ is given by the diagram in Fig.~\ref{fig:semilept} and is written as
\begin{eqnarray}
&&
\left\langle B_{(c)} (p_2) 
\left|\bar{q} O_{\mu} b \right| \Upsilon(1S) (\epsilon_1 , p_1) \right\rangle 
=\epsilon_1^{\alpha} T^{VP}_{\mu\alpha}
\nn
T^{VP}_{\mu\alpha} &=& 3 g_{\Upsilon(1S)} g_P \int\!\! \frac{d^4k}{(2\pi)^4 i}
\widetilde\Phi_{\Upsilon(1S)}[-(k + w_{13}p_1)^2] 
\widetilde\Phi_P[-(k + w_{23}p_2)^2]
\nn
&\times& \Tr\left[S_2(k+p_2) O_{\mu} S_1(k+p_1)\gamma_\alpha S_3(k)
\gamma_5\right]\nn
&\equiv& \frac{1}{m_1 + m_2}
\left[- g_{\mu\alpha}pqA_0(q^2) + p_{\mu}p_{\alpha}A_+(q^2)+q_{\mu}p_{\alpha}
A_-(q^2) + i \varepsilon_{\mu\alpha\gamma\delta}p^{\gamma}q^{\delta}V(q^2)\right],
\label{eq:VP}
\end{eqnarray}
where $k$ is the loop momentum and $w_{ij}=m_{q_j}/(m_{q_i}+m_{q_j})$ $(i,j=1,2,3)$.  

The form factors are then calculated using standard one-loop calculation techniques (see, e.g. Ref.~\cite{Ivanov:2015tru}).  The main steps are as follows. First, one substitutes the Gaussian form for the vertex functions in Eq.~(\ref{eq:vertexf}) into Eq.~(\ref{eq:VP}). Second, one uses the Fock-Schwinger representation for the quark propagator
\begin{equation}
	S_{q_i} (k) = (m_{q_i} + \not\! k)\int\limits_0^\infty \!\!d\alpha_i\,e^{-\alpha_i (m_{q_i}^2-k^2)}.
	\label{eq:Fock}
\end{equation}
Third, one treats the integrals over the Fock-Schwinger parameters $0\le \alpha_i<\infty$ by introducing an additional integration which converts the set of 
these parameters into a simplex as follows
\begin{equation}
	\prod\limits_{i=1}^n\int\limits_0^{\infty} 
	\!\! d\alpha_i f(\alpha_1,\ldots,\alpha_n)
	=\int\limits_0^{\infty} \!\! dtt^{n-1}
	\prod\limits_{i=1}^n \int\!\!d\alpha_i 
	\delta\left(1-\sum\limits_{i=1}^n\alpha_i\right)
	f(t\alpha_1,\ldots,t\alpha_n).
	\label{eq:simplex}  
\end{equation}
Note that Feynman diagrams are calculated in the Euclidean region where $p^2=-p^2_E$. The vertex functions fall off in the Euclidean region, therefore no ultraviolet divergence appears. In order to avoid possible thresholds in the Feynman diagram, we introduce a universal infrared cutoff which effectively guarantees the confinement of quarks within hadrons
\begin{equation}
\int\limits_0^\infty dt (\ldots) \to \int\limits_0^{1/\lambda^2} dt (\ldots).
\label{eq:conf}
\end{equation}
Each form factor for the semileptonic transition $\Upsilon(1S)\to B_{(c)}$ is finally turned into a three-fold integrals of the general form
\begin{equation}
	F(q^2)
	=\int\limits_0^{1/\lambda^2} \!\! dtt^2
	\int\limits_0^1 \!\! d\alpha_1 \int\limits_0^{1-\alpha_1} \!\!  d\alpha_2 \,
		f(t\alpha_1,t\alpha_2).
	\label{eq:ff}  
\end{equation}
The expressions for $f(t\alpha_1,t\alpha_2)$ are obtained by a FORM code written by us. The numerical calculation of the three-fold integrals are done by using FORTRAN codes with the help of NAG library (see Appendix).

\section{Numerical results}
\label{sec:result}
Before listing our numerical results, we briefly discuss the estimation of the theoretical errors in our approach. It should be reminded that all phenomenological quark models of hadrons are simplified physics picture, and therefore it is very difficult to treat the theoretical error rigorously. The main source of uncertainties come from the free parameters in Table~\ref{tab:para}. They are obtained by a least-squares fit of leptonic and electromagnetic decay constants to experimental data and/or Lattice QCD. The allowed deviation in the fit is in the range 5--10\%. This range can be used as reasonable estimation of the model's errors. Moreover, the CCQM has been applied to study a broad range of hadron decay processes. We observed that our predictions often agree with experimental data within 10\%. Therefore, we estimate the theoretical error of the predictions in this paper to be about 10\%.

%%%% begin adding
Note that the 10\% error arises from the fact that the model contains free parameters, which were determined through a fit to the experimental data (see Ref.~\cite{Ivanov:2011aa}). These fitted parameters carry their own uncertainties, and these uncertainties propagate through to the form factors and the decay widths. As a result, the overall error remains below 10\%. Regarding the errors associated with the numerical calculation of the multifold integrals, they are negligible, as we used a highly reliable subroutine from the NAG library. 
This ensures that the precision of the numerical integration does not contribute significantly to the overall error. More details on the NAG subroutine is given in the Appendix. 
%%%% end adding
\subsection{Form factors}
\begin{figure}[htbp]
	\begin{tabular}{lr}
		\includegraphics[width=0.50\textwidth]{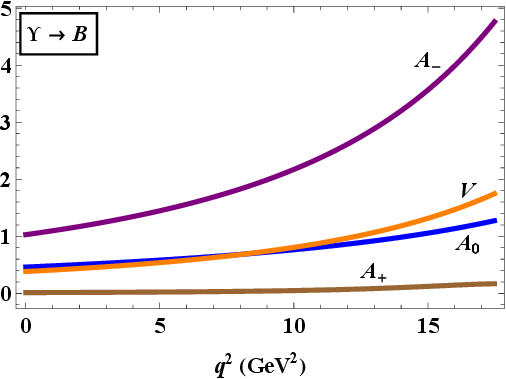} &
		\includegraphics[width=0.50\textwidth]{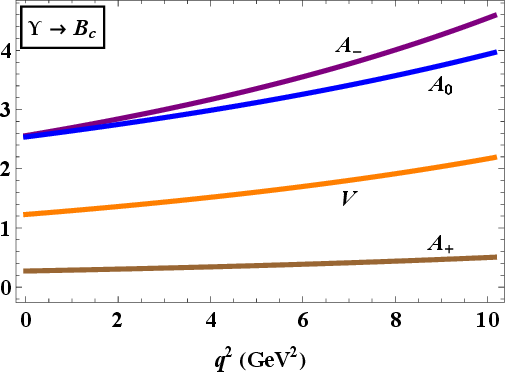}
	\end{tabular}
	\caption{Our results for the form factors of the $\Upsilon(1S) \to B$ (left) and 
		$\Upsilon(1S) \to B_c$ (right) transitions.} 
	\label{fig:FF}
\end{figure}

%%%
In Fig.~\ref{fig:FF} we present the form factors of the $\Upsilon(1S) \to B_{(c)}$ transitions 
in the full range of momentum transfer 
$0\le q^2 \le q^2_{\rm max} = (m_{\Upsilon(1S)}-m_{B_{(c)}})^2$. It is worth mentioning that in the CCQM, the form factors are directly calculated in the whole physical range without any extrapolation as usually seen in Lattice QCD and QCD Sum Rules. We then parametrize the $q^2$ dependence of the form factors by using a general dipole approximation
 \begin{equation}
 F(q^2)=\frac{F(0)}{1 - a s + b s^2}, \quad s\equiv\frac{q^2}{m^2_{\Upsilon(1S)}}.
 \label{eq:dipoleFF}
 \end{equation}
The dipole-approximation parameters for the  $\Upsilon(1S) \to B_{(c)}$ form factors are displayed in Table~\ref{tab:FF}. We also list here the values of the form factors at zero recoil, i.e., at $q^2_{\rm max}$. We compare the form factors at maximum recoil $(q^2 = 0)$ with other theoretical studies in Table~\ref{tab:FFq0}.
%
%%%%%%%
\begin{table}[ht]
	\caption{Parameters of the dipole approximation for $\Upsilon(1S)\to B_{(c)}$ form factors and the form-factor values at zero recoil $q^2_{\rm max}$.}
	\renewcommand{\arraystretch}{0.7}
\begin{ruledtabular}
		\begin{tabular}{c|cccc|cccc}
			\multicolumn{1}{c|}{} &\multicolumn{4}{c|}{$\Upsilon(1S)\to B$} 
			&\multicolumn{4}{c}{$\Upsilon(1S)\to B_c$} \\
			\hline
			& $ A_0 $ & $  A_+  $ & $  A_-  $ & $  V  $ 
			& $ A_0 $ & $  A_+  $ & $  A_-  $ & $  V  $ 
			\\
			\hline
	$F(0)$ & 0.46 & 0.013 & 1.03 & 0.38 & 2.54 & 0.27 & 2.56 & 1.23
			\\
	$a$    & 3.93 & 8.99  & 5.64 & 5.63 & 3.49 & 4.91 & 4.57 & 4.54
			\\
	$b$    & 3.41 & 21.9  & 8.34 & 8.37 & 2.83 & 7.43 & 5.93 & 5.89
			\\ 
	$F(q^2_{\rm max})$ 
		   & 1.28 & 0.18 & 4.78 & 1.75 & 3.96 & 0.50 & 4.59 & 2.19
		\end{tabular}
\end{ruledtabular}
		\label{tab:FF}
\end{table}
%%%%%%%%%%%
%%%%%%%
\begin{table}[ht]
	\caption{Form factors at maximum recoil $q^2=0$ in the CCQM and other theoretical studies. The values in the row BS were obtained based on the form-factor graphs of Ref.~\cite{Wang:2016dkd} and private communication with its authors.}
	\renewcommand{\arraystretch}{0.7}
	\begin{ruledtabular}
		\begin{tabular}{c|cccc|cccc}
			\multicolumn{1}{c|}{} &\multicolumn{4}{c|}{$\Upsilon(1S)\to B$} 
			&\multicolumn{4}{c}{$\Upsilon(1S)\to B_c$} \\
			\hline
			& $ A_0(0) $ & $  A_+(0)  $ & $  A_-(0)  $ & $  V(0)  $ 
			& $ A_0(0) $ & $  A_+(0)  $ & $  A_-(0)  $ & $  V(0)  $ 
			\\
			\hline
			This work & 0.46 & 0.013 & 1.03 & 0.38 & 2.54 & 0.27 & 2.56 & 1.23
			\\		
			BSW~\cite{Dhir:2009rb} & & & & & $3.06^{+0.10}_{-0.05}$ & $0.38^{+0.06}_{-0.09}$ & & $1.61^{+0.01}_{-0.01}$
			\\
			Ref.~\cite{Sharma:1998gc} &   &   &   &   & 4.99 & 1.01 & & 1.01 \\
			NRQCD~\cite{Chang:2016gyw} &   &   &   &   & 3.46 & 0.51 & & 1.66 \\
			BS~\cite{Wang:2016dkd} & 0.20 & 0.031 & 0.29 & 0.15  & 1.63 & 0.26 & 1.28 & 0.80  
			\\
		\end{tabular}
	\end{ruledtabular}
	\label{tab:FFq0}
\end{table}
%%%%%%%%%%%
\subsection{Branching fractions}
We present our results for the branching fractions of the semileptonic decays $\Upsilon(1S)\to B_{(c)}\ell\bar{\nu}_\ell$ ($\ell = e,\mu,\tau$) in Table~\ref{tab:BF}. We also show in this table the  relevant predictions of other theoretical studies for comparison. Our predictions agree well with the Bethe-Salpeter--approach results~\cite{Wang:2016dkd}. Regarding the results obtained using the Bauer-Stech-Wirbel model~\cite{Dhir:2009rb}, the branching fractions for the electron and muon modes in this study agree with ours, but the one for the tau mode disagrees. The results obtained in NRQCD~\cite{Chang:2016gyw} for the branching fractions can be seen to be larger than those in other studies including ours.
%%%%% 

\begin{table}[ht]
	\caption{Branching fractions of $\Upsilon(1S)$ semileptonic decays in the CCQM, the Bethe-Salpeter (BS) approach~\cite{Wang:2016dkd}, the Bauer-Stech-Wirbel (BSW) model~\cite{Dhir:2009rb}, and the NRQCD framework~\cite{Chang:2016gyw}.}
	\label{tab:BF}
	\begin{center}
			\begin{tabular}{c|c|c|c|c|c}
				\hline\hline
			Channel  & Unit & This work  &  BS~\cite{Wang:2016dkd} & BSW~\cite{Dhir:2009rb} & NRQCD~\cite{Chang:2016gyw}\\
			\hline
			$\Upsilon(1S)\to B e \bar\nu_e$ & $10^{-13}$ & 5.96 & $7.83^{+1.40}_{-1.20}$ & &\\
			$\Upsilon(1S)\to B \mu \bar\nu_\mu$ & $10^{-13}$ & 5.95 & $7.82^{+1.40}_{-1.20}$ & &\\
			$\Upsilon(1S)\to B \tau \bar\nu_\tau$ & $10^{-13}$ & 3.30 & $5.04^{+0.92}_{-0.79}$ & &\\
			$\Upsilon(1S)\to B_c e \bar\nu_e$ & $10^{-10}$ & 1.84 & $1.37^{+0.22}_{-0.19}$ & $1.70^{+0.03}_{-0.02}$ & $5.58^{+3.32\,\,+0.14\,\,+0.08}_{-1.54\,\,-0.12\,\,-0.18}$\\
			$\Upsilon(1S)\to B_c \mu \bar\nu_\mu$ & $10^{-10}$ & 1.83 & $1.37^{+0.22}_{-0.19}$ & $1.69^{+0.04}_{-0.02}$ & $5.58^{+3.32\,\,+0.14\,\,+0.08}_{-1.54\,\,-0.12\,\,-0.18}$\\
			$\Upsilon(1S)\to B_c \tau \bar\nu_\tau$ & $10^{-11}$ & 4.74 & $4.17^{+0.58}_{-0.52}$ & $2.90^{+0.05}_{-0.02}$ & $13.0^{+7.7\,\,+0.3\,\,+0.2}_{-3.5\,\,-0.3\,\,-0.4}$\\
			\hline\hline
		\end{tabular}
	\end{center}
\end{table} 
%%%
It is interesting to consider the ratio 
$R_{B_{(c)}}\equiv\mathcal{B}(\Upsilon(1S)\to B_{(c)} \tau \bar{\nu}_\tau)/\mathcal{B}(\Upsilon(1S)\to B_{(c)} \ell \bar{\nu}_\ell)$, $(\ell = e,\mu)$, where a large part of theoretical and
experimental uncertainties cancels. 
%%%% begin adding
Note that the key advantage in considering this ratio is the cancellation of various uncertainties. Specifically, the hadronic form factors, which describe the $ \Upsilon(1S) \to B_c $ transition and carry theoretical uncertainties, largely cancel out. Similarly, uncertainties from radiative corrections, including QCD, QED, and electroweak interactions, also cancel due to their similar impact on both decay channels in the ratio. Additionally, CKM-matrix-element uncertainties are eliminated, as they affect both processes identically. These cancellations make the ratio a particularly clean observable for studying potential deviations from the SM and probing NP effects.
%%%% end adding

We list in~(\ref{eq:RB}) and~(\ref{eq:RBc}) all available 
predictions for $R_{B_{(c)}}$ up till now:
\begin{equation}
	R_B\equiv \frac{\mathcal{B}(\Upsilon(1S) \to B \tau \bar{\nu})}
	{\mathcal{B}(\Upsilon(1S) \to B \ell \bar{\nu})}
	=\left\{\begin{array}{lc}
		0.64 \qquad & \qquad \text{{BS}~\cite{Wang:2016dkd}}  \\
		0.55  \qquad & \qquad \text{This work}
	\end{array}\right.
	,
	\label{eq:RB}
\end{equation}
%%%%
\begin{equation}
R_{B_c}\equiv \frac{\mathcal{B}(\Upsilon(1S) \to B_c \tau \bar{\nu})}
{\mathcal{B}(\Upsilon(1S) \to B_c \ell \bar{\nu})}
=\left\{\begin{array}{lc}
	0.30 \qquad & \qquad \text{{BS}~\cite{Wang:2016dkd}} \\
	0.17 \qquad & \qquad \text{{BSW}~\cite{Dhir:2009rb}}  \\
	0.24 \qquad & \qquad \text{{NRQCD}~\cite{Chang:2016gyw}}\\
	0.26  \qquad & \qquad \text{This work}
\end{array}\right.
.
\label{eq:RBc}
\end{equation}
Our results for the ratios $R_{B}$ and $R_{B_c}$ agree well with those in the BS approach. Meanwhile, the result for $R_{B_c}$ in the BSW is about two times smaller than the BS and our predictions. Moreover, NRQCD prediction for $R_{B_c}$ is very close to ours. Therefore, we propose that the value $R_{B_c}\simeq 0.3$ is a reliable prediction. 
%%%%%%
\subsection{\boldmath{$\Upsilon(1S)\to B_{c}\ell\bar{\nu}_\ell$} beyond the Standard Model}
%%%%%%
As already mentioned in Sec.~\ref{sec:intro}, the semileptonic decay $\Upsilon(1S)\to B_{c}\ell\bar{\nu}_\ell$ is induced by the quark-level transition $b\to c\ell\bar{\nu}_\ell$ and can be linked with the $R_D^{(*)}$ anomaly. It is therefore interesting to probe the possible New Physics (NP) effects in the $\Upsilon(1S)$ semitauonic decay. Based on the current status of the anomalies, we assume that NP only affects leptons of the third generation and modify the effective Hamiltonian for the quark-level transition 
$b \to c \tau^- \bar{\nu}_{\tau}$ as follows 
\begin{eqnarray}
{\mathcal H}_{eff} &=&
2\sqrt {2}G_F V_{cb}[(1+V_L)\mathcal{O}_{V_L}+V_R\mathcal{O}_{V_R}],
\label{eq:Heff}
\end{eqnarray}
where the four-fermion operators are written as
\begin{eqnarray}
\mathcal{O}_{V_L} = 
\left(\bar{c}\gamma^{\mu}P_Lb\right)\left(\bar{\tau}\gamma_{\mu}P_L\nu_{\tau}
\right),
\nn
\mathcal{O}_{V_R} =
\left(\bar{c}\gamma^{\mu}P_Rb\right)
\left(\bar{\tau}\gamma_{\mu}P_L\nu_{\tau}\right).
\label{eq:operators}
\end{eqnarray}
Here, $P_{L,R}=(1\mp\gamma_5)/2$ are the left and right projection operators, and 
$V_{L,R}$ are the complex Wilson coefficients governing 
the NP contributions. In the SM one has $V_{L,R}=0$. 

The invariant matrix element of the semileptonic decay 
$\Upsilon(1S)\to B_{c}\tau\bar{\nu}_\tau$ is then written as
\begin{eqnarray}
\mathcal{M}|_{\rm NP}&=&
\frac{G_FV_{cb}}{\sqrt{2}}\Big[
(1+V_R+V_L)\langle B_c|\bar{c}\gamma^\mu b|\Upsilon(1S)\rangle 
\bar{\tau}\gamma_\mu(1-\gamma^5)\nu_\tau\nn
&&
+(V_R-V_L)\langle B_c|\bar{c}\gamma^\mu\gamma^5b|\Upsilon(1S)\rangle
\bar{\tau}\gamma_\mu(1-\gamma^5)\nu_\tau\Big].
\label{eq:amplitude-full}
\end{eqnarray}
Note that the axial hadronic currents do not contribute to the  $P\to P^\prime$ transition. Therefore, assuming that NP appears in both $\bar{B}^0\to D$ and $\bar{B}^0\to D^*$ transitions, the case of pure $V_R-V_L$ coupling is ruled out. The branching fraction $\mathcal{B}(\Upsilon(1S)\to B_{c}\tau\bar{\nu}_\tau)$ is therefore modified according to
\begin{eqnarray}
	\mathcal{B}(\Upsilon(1S)\to B_{c}\tau\bar{\nu}_\tau)|_{\rm NP} 
	=|1+V_L+V_R|^2\mathcal{B}(\Upsilon(1S)\to B_{c}\tau\bar{\nu}_\tau)|_{\rm SM}.
\end{eqnarray}

\begin{figure}[ht]
	\centering
	\begin{tabular}{cc}
		\hspace{-1cm}
		\includegraphics[width=0.50\textwidth]{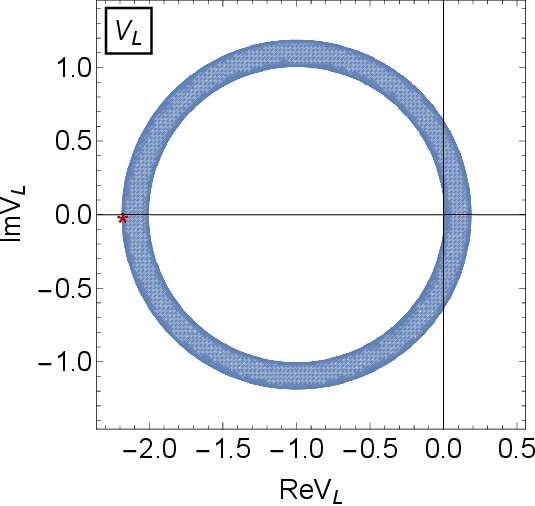}&
		\includegraphics[width=0.50\textwidth]{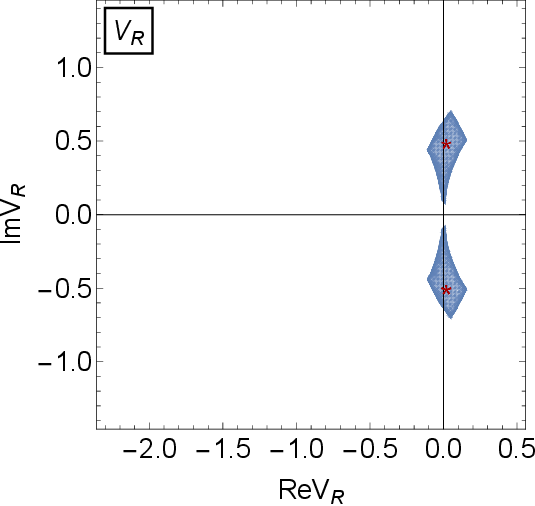}
	\end{tabular}
	\caption{Constraints on the complex Wilson coefficients $V_L$ and $V_R$ from the measurements of $R_D$, $R_{D^*}$, $R_{J/\psi}$, and $F_L^{D^*}({B} \to D^{\ast} \tau\nu_\tau)$ within $2\sigma$, and from the branching fraction ${\cal B}(B_c\to\tau\nu_\tau)\leq 10\%$. The allowed regions are indicated in light-blue color. The asterisk symbols indicate the best-fit values.}
	\label{fig:2sigma}
\end{figure}

By assuming the dominance of only one NP operator at a time, the allowed regions for the NP Wilson coefficients $V_{L(R)}$ are obtained using experimental data for the ratios of branching fractions $R_D=0.344\pm 0.026$, $R_{D^*}=0.285\pm 0.012$~\cite{HFLAV:2022esi}, $R_{J/\psi}=0.71\pm 0.17\pm 0.18$~\cite{LHCb:2017vlu}, the upper limit ${\cal B}(B_c\to \tau\nu) \leq 10\,\%$ from the LEP1 data~\cite{Akeroyd:2017mhr}, and the longitudinal polarization fraction of the $D^*$ meson $F_L^{D^*}({B} \to D^{\ast} \tau\bar{\nu}_\tau)=0.43\pm 0.06\pm 0.03$~\cite{LHCb:2023ssl}. The relevant form factors for the transitions $B\to D^{(*)}$ and $B_c\to J/\psi$ were calculated in our paper~\cite{Tran:2018kuv}. In Fig.~\ref{fig:2sigma}, we show the allowed regions for $V_L$ and $V_R$ within $2\sigma$. In each region, we find a best-fit value and mark it with an asterisk.
%%%%

We summarize our predictions for the branching fraction $\mathcal{B}(\Upsilon(1S)\to B_c \tau \bar{\nu}_\tau)$ and the ratio of branching fractions $R_{B_c}$ in Table~\ref{tab:ratio}. The row labeled by SM (CCQM) contains our predictions within the SM with the CCQM form factors. The predicted intervals for the observables in the presence of NP are given in correspondence with the $2\sigma$ allowed regions of the NP Wilson coefficients depicted in Fig.~\ref{fig:2sigma}. It is worth mentioning that the $V_L$ NP scenario can enhance the physical observables by a factor of 6.
%%%%%%%%%%%%%%%
\begin{table}[htbp] 
	\vspace{-3mm}
	\begin{center}
		\caption{Observables in the SM and in the presence of NP.}
		\vspace{2mm}
		\label{tab:ratio}
		\begin{tabular}{|c|c|c|c|}
			\hline
			Quantity &\quad  SM (CCQM) \qquad 
			&\quad  $V_R$ \qquad   
			&\quad  $V_L$ \qquad  
			\\
			\hline
			$10^{11}\times\mathcal{B}(\Upsilon(1S)\to B_c \tau \bar{\nu}_\tau)$
			&\quad 4.74 \quad
			&\quad $(4.77,7.07)$\quad & \quad $(4.74,27.3)$ \quad
			\\
			\hline
	   		$R_{B_c}$
			&\quad 0.26\quad
			&\quad $(0.26,0.39)$\quad &\quad $(0.26,1.50)$ \quad
						\\
					\hline
		\end{tabular}
	\end{center}
	\vspace{-5mm}
\end{table}
%%%%%%

%%%%% begin adding
Finally, it is worth mentioning the possibility of experimental observation of the semileptonic decays $\Upsilon(1S)\to B_{(c)}\ell\bar{\nu}_\ell$. As already pointed out in Ref.~\cite{Chang:2016gyw}, a branching fraction at order $10^{-10}$ of the decays $\Upsilon(1S)\to B_c\ell\bar{\nu}_\ell$ is almost impossible to reach at LHCb, CMS, and Belle-II. Even though the accumulated number of $\Upsilon(1S)$ samples at these colliders can be as large as $10^{10}$--$10^{11}$, the low reconstruction efficiency (assumed to be several percents) makes it extremely difficult to observe any significant signal. Even if NP effects enhance the branching fractions by an order of magnitude (in the case of semitauonic decay), it is still a challenging task to study these decays experimentally (see Ref.~\cite{Chang:2016gyw} for more details). 
%%%%% end adding

\section{Summary}
\label{sec:sum}
This paper represents a new study of the semileptonic decays $\Upsilon(1S)\to B_{(c)}\ell\bar{\nu}_\ell$, where $\ell = e,\mu,\tau,$ inspired by the recent search for similar rare weak decays of $J/\psi$ at BESIII. The relevant form factors for the $\Upsilon(1S)\to B_{(c)}$ transitions are calculated in the whole momentum transfer squared region in the framework of the Covariant Confined Quark Model. Predictions for the branching fractions and their ratios are reported and compared to other theoretical studies. A good agreement with the results of the Bethe-Salpeter approach was found. However, our prediction for the ratio of branching fractions $R_{B_{(c)}}\equiv\mathcal{B}(\Upsilon(1S)\to B_{(c)} \tau \bar{\nu}_\tau/\mathcal{B}(\Upsilon(1S)\to B_{(c)} \ell \bar{\nu}_\ell)$ disagrees with the Bauer-Stech-Wirbel model prediction. We predict $R_{B_c}=0.26$ and $R_{B}=0.55$ which are close to the values $R_{B_c}=0.30$ and and $R_{B}=0.64$ obtained in the Bethe-Salpeter approach. We also extend the SM effective Hamiltonian for the $b\to c\tau\bar{\nu}_\tau$ transition by including left- and right-handed 4-fermion operators of dimension six. The relevant Wilson coefficients are obtained based on experimental data. Using the $2\sigma$ allowed regions for these coefficients, we found that the branching fraction of the tau mode as well as the ratio $R_{B_c}$ can be enhanced by about an order of magnitude. There have been only few theoretical calculations for $\Upsilon(1S)$ semileptonic decays to date. This study therefore provides more insights for experimental test of the SM, as well as the search for NP at future colliders.

\begin{acknowledgments}
C.~T.~T. and H.~C.~T. thank HCMC University of Technology and Education for support in their work and scientific collaboration. This work is supported by Ho Chi Minh City University of Technology and Education under Grant T2023-76.
\end{acknowledgments}

\appendix
\section{NAG subroutine d01fcf} 

For numerical calculations of multifold integrals we have used
the NAG subroutine d01fcf. For instance, in the case of threefold
integrals it works as follows:\\

\noindent 
external FA0: function supplied by the user (our form factor)\\
eps=$10^{-5}$: the relative errors acceptable by the user \\
maxpts=10000000:  the maximum number of integrand evaluations \\
minpts=0: the minimum number of integrand evaluations \\
lenwrk=500000  \\
wrkstr(500000)  \\
ifail=0  \\
ndim =3: the number of dimensions \\
finval: contains the best estimate obtained for the integral \\
acc:  contains the estimated relative error in finval \\
\noindent

do 20 k=1,3\\
a(k)=0.d0 \\
b(k)=1.d0\\

20     continue\\

\noindent
call d01fcf(ndim,a,b,minpts,maxpts,FA0,eps,acc,lenwrk,wrkstr,finval,ifail)\\

A0=finval\\
acc=$10^{-4}$. \\

\end{document}